\documentclass[a4paper]{jpconf}
\usepackage{graphicx}
\begin{document}
\title{Neutrinos and Future Concordance Cosmologies}

\author{Peter Adshead and Richard Easther}

\address{Department of Physics, Yale University, New Haven, CT 06520, USA}

\ead{richard.easther@yale.edu}

\begin{abstract}
We review the free parameters in the concordance cosmology, and those
which might be added to this set as the quality of astrophysical data
improves. Most concordance parameters encode information about
otherwise unexplored aspects of high energy physics, up to the GUT
scale via the ``inflationary sector,'' and possibly even the Planck
scale in the case of dark energy.  We explain how neutrino properties
may be constrained by future astrophysical measurements. Conversely,
future neutrino physics experiments which directly measure these
parameters will remove uncertainty from fits to astrophysical data,
and improve our ability to determine the global properties of our
universe. 
\end{abstract}

\section{Introduction}

Over the last decade, a ``concordance cosmology'' has emerged, which is consistent with all major astrophysical datasets and offers a description of the overall history and global dynamics of our  universe.\footnote{This is a qualitative discussion of cosmological neutrino constraints, based on a paper given by RE at the Neutrino 2008 meeting. For an introduction to the technical literature see e.g. \cite{Komatsu:2008hk,Lesgourgues:2006nd} and references therein. }. The  current ``benchmark'' parameter set is provided by the WMAP team's analysis of their 5-year dataset, in conjunction with other survey information \cite{Komatsu:2008hk}.       The concordance cosmology  contains just six free parameters, summarized in Table~\ref{table:conc}.  Most of these  parameters are fixed by unknown fundamental processes, and thus encode information about ``new physics.''    Looking at Table~\ref{table:conc},  three parameters describe  the present day energy-density of our universe: baryonic matter, dark matter, and dark energy.  Baryonic matter is of course familiar, but the {\em quantity\/} of baryonic matter in the universe (relative to the photon number density) is not yet predicted by fundamental theory.    There is no non-astrophysical evidence for  dark matter or dark energy, although the concordance cosmology implies that the {\em dark sector\/} comprises $\sim 95\%$ of the current energy density, dominating both the dynamics of large scale structure and the overall expansion of the universe.\footnote{One might replace dark matter or dark energy by either modifying gravity on very large scales, or (in the case of dark energy) allowing the underlying spacetime geometry to differ from the homogeneous and isotropic background of the concordance model. Such proposals certainly exist, and -- like the concordance model -- will be subject to increasingly stringent tests as the data improves.}   It is not unreasonable to expect that the properties of the (non-baryonic) dark matter  involve LHC-scale particle physics, whereas understanding dark energy could shed light on  quantum gravity.  

 Of the remaining concordance  parameters, $h$ reflects the current expansion rate of the universe. When the first stars turn on, the universe reinonizes at a time parametrized by the optical depth $\tau$.  Neither of these quantities  relies on ``new physics.''  However, the primordial perturbation spectrum is widely believed to have been fixed during the inflationary phase, which can occur anywhere between the GUT and TeV scales. In this case,  $A_s$ and $n_s$ are determined by the properties of very high energy particle physics. 
The standard concordance cosmology assumes only the three known neutrino species. The mass differences between these species are well established \cite{Yao:2006px}.  According to the well-understood thermal history of the hot big bang, the neutrino sector decouples  {\em before\/} the photon gas cools until  it no longer pair-produce electrons and positrons. The electrons and positrons then annihilate electromagnetically, heating the photons relative to the neutrino populations.   Taking into account the Fermi statistics of the neutrinos, one computes
\begin{equation}
T_\nu= \left(\frac{4}{11}\right)^{1/3} T_\gamma \, ,
\end{equation}
from which one also deduces their number density (e.g. \cite{Kolb:1990vq}).

\begin{table}[tb]
\begin{center}
\begin{tabular}{||c|l|c | c ||}
\hline \hline
\bf Label & \bf Definition & \bf Physical Origin & \bf Value \\
\hline
$\Omega_b$ & Baryon Fraction & Baryogenesis &  $0.0462 \pm 0.0015$\\
 \hline
 $\Omega_{\rm CDM}$ &Dark Matter Fraction & TeV Scale Physics (?) & $0.233 \pm 0.013$ \\
 \hline
 $\Omega_{\Lambda}$ & Cosmological Constant & Unknown  & $0.721 \pm 0.015$ \\
  \hline \hline
  $\tau $ & Optical Depth  & First Stars & $0.084 \pm 0.016$ \\
  \hline \hline
 $h$ & Hubble Parameter & Cosmological Epoch & $0.701 \pm 0.013$\\
 \hline   \hline
 $A_s$ & Perturbation Amplitude  & Inflation & $(2.45 \pm 0.09) \times 10^{-9}$ \\
 \hline
 $n_s$ & Perturbation Spectrum & Inflation & $0.960 \pm 0.014$ \\
 \hline \hline
\end{tabular}
\caption{ \label{table:conc}  The parameters of the current cocordance cosmology.  The contribution of the $i$-th constituent of the overall energy density is measured by $\Omega_i$, in units where the critical density corresponding to a spatially flat universe is unity.  All current observations are consistent with flatness, so  $\Omega_b + \Omega_{\rm CDM} + \Omega_\Lambda \equiv 1$, and there are thus just six free parameters.  ``Spectrum'' refers to the primordial scalar or density perturbations, parameterized by equation~\ref{eqn:spec}. }
\end{center}
\end{table}

\section{Neutrinos and Concordance Cosmologies}

\begin{table}[t]
\begin{center}
\begin{tabular}{||c|l|c||}
\hline \hline
\bf Label & \bf Definition & \bf Physical Origin \\
\hline
 $\Omega_k$ & Curvature & Initial Conditions \\
\hline
 $\Sigma m_\nu$ & Neutrino Mass & Beyond-SM Physics \\
\hline
 $N_\nu$ & Neutrino-like Species & Beyond-SM Physics \\
\hline
 $w $ & Dark Energy Equation of State& Unknown \\
 \hline \hline
 $Y_{\rm He}$ & Helium Fraction & Nucleosynthesis \\
\hline \hline
 $\alpha_s$ &Spectral ``Running''&   Inflation  \\
\hline
 $A_t$ &Tensor Amplitude &   Inflation  \\
\hline
 $n_t$ &Tensor Spectrum &   Inflation  \\
\hline
 $f_{\rm NL}$ & Non-Gaussianity & Inflation (?) \\
\hline
 $S$ & Isocurvature & Inflation \\ 
\hline \hline
\end{tabular}
\caption{ \label{table:concfuture} Parameters in possible future concordance cosmologies.    The tensor or gravity wave spectrum  is parametrized as $A_t
(k/k_\star)^{n_t}$ while $w$ and $\alpha_s$, could be extended to dark energy with a non-trivial  equation of
state ($w'$), or a spectrum with  ``features'' and more generally a scale-dependence that cannot be
parameterized by a spectral index and its running alone.  In many inflationary models, $A_t$ and $n_t$ are correlated and replaced by a single parameter, $r\sim A_t/A_s$, with $n_{t} = -r/8$.}
\end{center}
\end{table}

The parameter set of the concordance cosmology  is chosen to maximize the $\chi^2$ per degree of freedom, or via Bayesian evidence (e.g. \cite{Trotta:2008qt}).   One can set upper limits on many other parameters, and  these would be added to the ``concordance'' set if they were detected in the data  \cite{Liddle:2004nh}. Table \ref{table:concfuture} lists some of the most commonly discussed parameters. Inflation is assumed to set  the height and spectral index of the power spectrum, which is parametrized by
\begin{equation} \label{eqn:spec}
P_s(k) = A_s(k_\star) \left(\frac{k}{k_\star}\right)^{n_s(k_\star)-1+ \frac{1}{2} \alpha_s(k_\star) \ln (k/k_\star)}\, 
\end{equation}
where $k_\star$ is a specified by otherwise irrelevant pivot scale. The constraint on $n_s$  given in Table~\ref{table:conc} is much weaker than it might appear, as $n_s=1$ corresponds to the ``default'' case of a scale-free or Harrison-Zel'dovich spectrum.   In practice, $A_s$ is a free parameter in  most inflationary models, while $n_s$ and the running $\alpha$ are fixed by the detailed physics of the inflationary era.  Taken together with possible measurements of a primordial tensor spectrum or non-Gaussianity, future concordance models could possess an ``inflationary sector''  containing several parameters,  which would  potentially shed light on GUT scale physics.

The concordance model assumes  a) three neutrino species, b) with zero mass and c)  relic abundances predicted by the conventional thermal history for the universe.   
This last point may be the most robust, as it is checked independently by   nucleosynthesis, which is sensitive to departures from the standard model and occurs at temperatures relatively close to neutrino freeze-out.  However, cosmological data limits the number of additional neutrino-like species which freeze out with a substantial population in the early universe.   
   As the universe expands, the relative {\em number\/} density of photons, neutrinos and massive particles is essentially constant. However, the energy of the photons and relativistic neutrinos redshifts, decreasing linearly with the expansion of the universe. Consequently, while the universe is initially radiation dominated, the contributions of radiation and matter cross over at the point of matter-radiation equality.  The current temperature of the microwave background is 2.725 K; from astrophysical data we can deduce that  matter-radiation equality occurs at redshift of around 3,200, and thus a temperature of 8,700 K.  At a redshift of slightly less than 1,100 (or T = 3,000 K) the universe recombines, and becomes transparent.\footnote{A temperature of 3,000 K corresponds to an energy $\sim$ 0.25 eV, which is less than the 13.6 eV required to ionize hydrogen. The universe contains approximately a billion photons for every proton, and cools well below 13.6 eV before the high energy tail of the blackbody distribution of photons cannot ionize hydrogen.}

 \begin{figure}[tb]
 \begin{center}
\includegraphics[width=18pc]{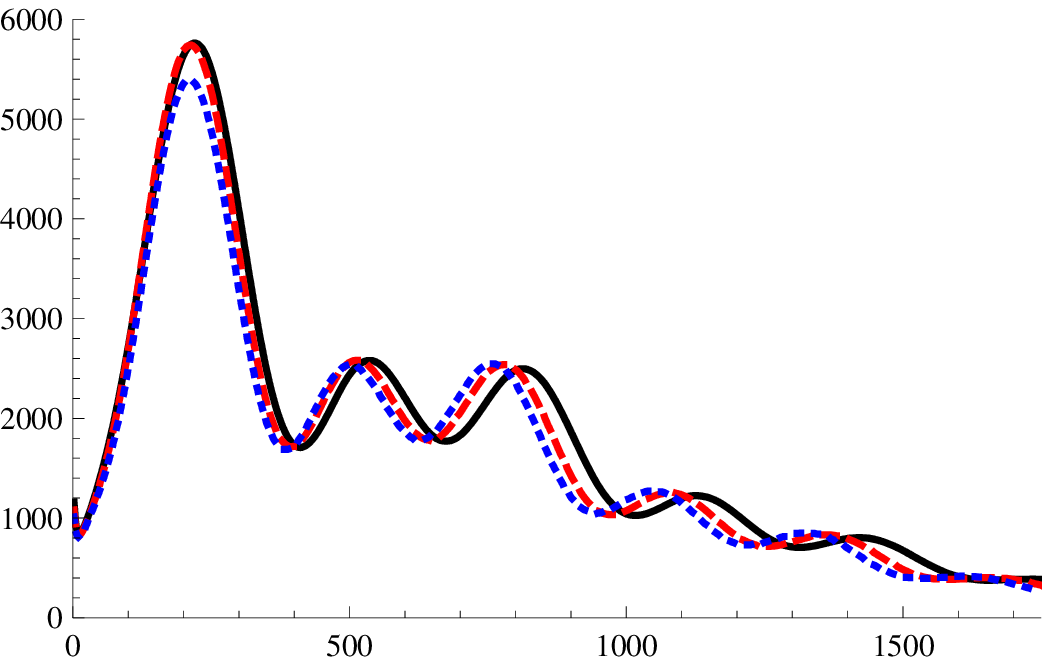}
\includegraphics[width=18pc]{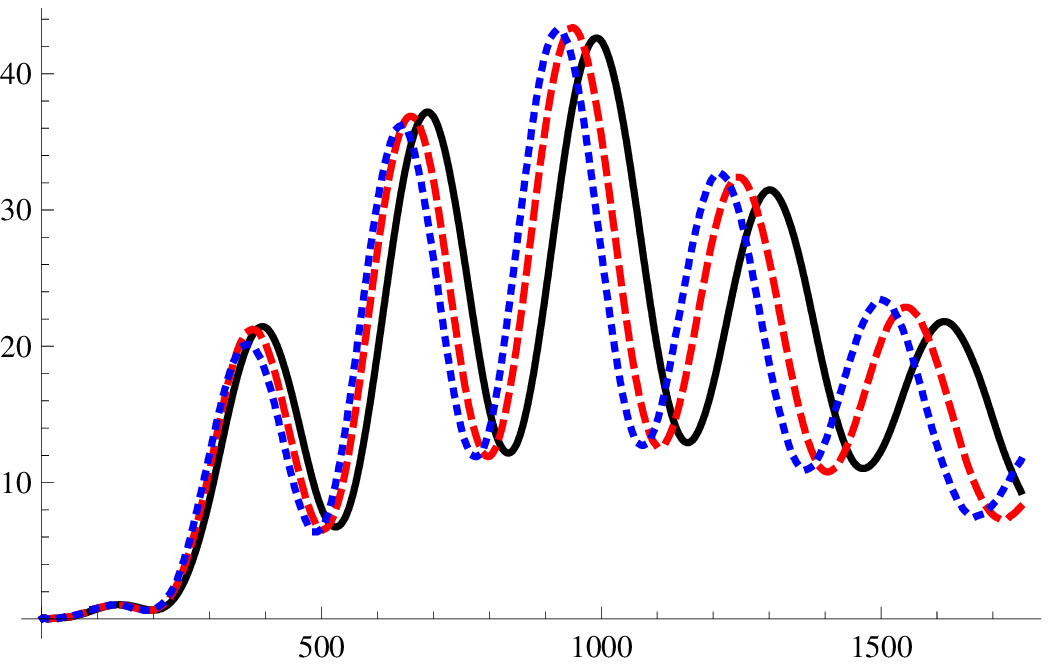}
\end{center}
\caption{\label{fig:camb}  We plot the CMB temperature ($\langle TT\rangle$, left) and the E-mode polarization   ($\langle EE\rangle$, right) power spectra, computed with the central values parameter values found by WMAP 5, and three neutrino species and $\Sigma m_\nu$ of 0 (solid), 2 (dashed) and 4 (dotted) eV respectively.  In the latter two cases, the neutrinos are massive at recombination, and the  peaks move to lower multipoles, while the relative height of the first peak begins to decrease sharply with $\Sigma m_\nu$.   }
\end{figure}
  
\begin{figure}[tb]
 \begin{center}
\includegraphics[width=36pc]{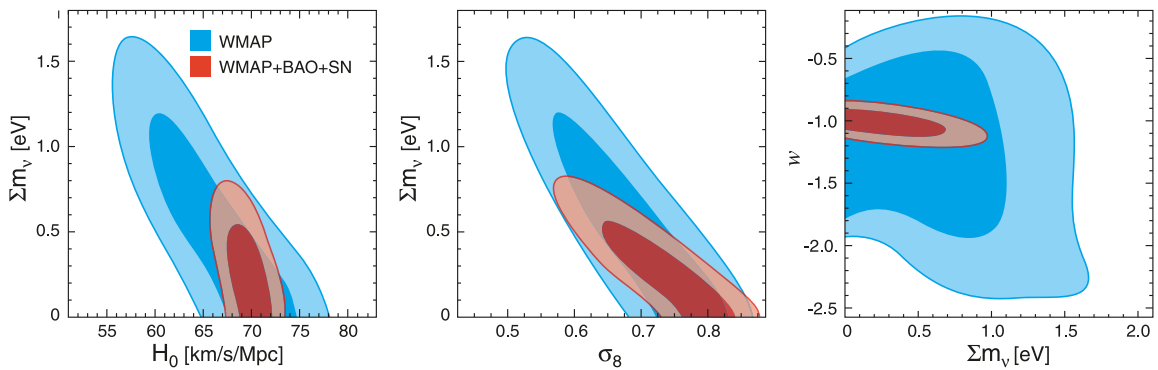}
\end{center}
\caption{\label{fig:data} WMAP5 results for the overall neutrino mass, where both $\Sigma m_\nu$ and the dark energy equation of state is allowed to vary. $\sigma_8$ is a derived parameter that quantifies the scale at which galaxy clusters lead to nonlinear density perturbations, and independent measurements of this parameter thus tighten constraints on $\Sigma m_\nu$.} 
\end{figure}

While neutrinos obviously do not interact directly with matter, they make their presence felt in two ways.  Firstly,   the relic neutrino background makes a nontrivial contribution to the expansion rate of the universe during recombination. Secondly,  a neutrino species with a rest mass significantly greater  than $0.25$ eV moves non-relativistically during recombination,  altering the detailed physics   of the peaks in the microwave background power spectrum (e.g. \cite{Lesgourgues:2006nd}), as illustrated in Figure~\ref{fig:camb}.    
Given the quality of present data,  neutrino masses constraints are usually written in terms of $\Sigma m_\nu$, with the individual masses  degenerate. Results from the  WMAP 5 analysis are shown in Figure~\ref{fig:data}. We see that {\em current\/} cosmological data puts a tight upper bound on the total mass of the neutrino sector.\footnote{For comparison, KATRIN ({\tt http://www-ik.fzk.de/$\sim$katrin/}) is projected to be sensitive $\nu_e$ mass of $0.2$ eV, which is approximately one third of the {\em current} cosmological upper bound on $\Sigma m_\nu$.}
 Figure~\ref{fig:camb} shows that even a relatively small $\Sigma m_\nu$ produces a detectable shift in the CMB power spectrum, given that the peaks have now been accurately measured.   However, this shift can be compensated for by adjustments to other cosmological parameters, an example of the well-known degeneracy in a cosmological parameter measurement. In particular, by modifying $h$ or equivalently $H_0$ (the current value of the Hubble constant), we can significantly increase the allowed value of $\Sigma m_\nu$.  However, this degeneracy is broken by adding further cosmological data -- in this case baryon acoustic oscillation (BAO) and high redshift supernovae, which provide measurements of $H_0$ independent of assumptions about $\Sigma m_\nu$, leading to a much tighter overall fit.

\section{Future Concordance Cosmologies}

 \begin{figure}[tb]
 \begin{center}
\includegraphics[width=24pc]{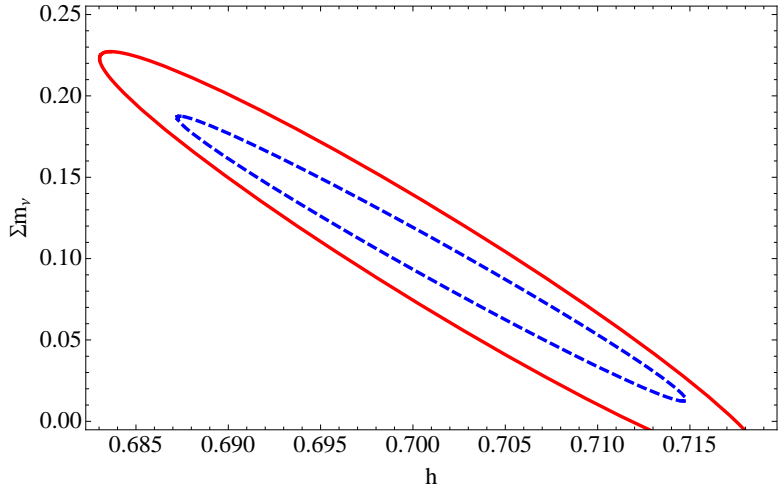}
\end{center}
\caption{\label{fig:forecast}  Forecast 1-$\sigma$ errors in the $\Sigma m_\nu$-$h$ plane, where the present day Hubble constant is $H_{0} =100\,h$ kms$^{-1}$Mpc$^{-1}$. The outer ellipse is derived from the Ideal satellite of \cite{Verde:2005ff,Adshead:2008vn} (roughly equivalent to ambitious CMBpol style proposals, e.g. \cite{Bock:2008ww}), for a concordance parameter set the usual $\Lambda$CDM variables, plus $r$ (the tensor-scalar ratio), $\alpha$ and $\Sigma m_\nu$.  The outer ellipse allows for finite signal/noise in the detectors, while the inner ellipse is the cosmic variance limit up to $\ell=1500$. Consequently, there is no guarantee CMB data alone will fix $\Sigma m_\nu$ Moreover, this forecast is optimistic as we have assumed that foregrounds are fully subtracted.}
\end{figure}

Oscillation data puts a lower bound of around 0.05 eV on $\Sigma m_\nu$, and the astrophysical bound is now within an order of magnitude of this limit.     Fisher matrix forecasts (see Figure 3)  show that a ``next generation'' CMB mission designed for excellent polarization  sensitivity might make a 1-$\sigma$ ``detection''  for  $\Sigma m_\nu = 0.1$ eV. The large degeneracy between $\Sigma m_\nu$ and $h$ is broken by data which constrains $h$ (or the $\sigma_8$ parameter). Future large scale structure and supernovae surveys  will greatly tighten current bounds on these parameters, so it is very likely that astrophysical measurements will eventually detect $\Sigma m_\nu$, and the concordance cosmology will contain at least one parameter related to the neutrino sector. More ambitiously, high redshift 21cm data may be sensitive to the individual neutrino masses, and thus reveal whether these fall into a regular or inverted hierarchy \cite{Pritchard:2008wy}. 

Unlike terrestrial experiments, cosmological neutrino constraints necessarily involve assumptions about the overall form of the universe, and their uncertainties are correlated with measurements of other cosmological parameters.  However, a further generation of terrestrial experiments may provide absolute determinations of at least one neutrino mass. In this case,  the concordance cosmological parameter set could once again contain no free parameters directly related to the neutrino sector.  By removing the freedom associated with the unknown neutrino masses when determining the overall form of the universe, experimental neutrino physics deepens our understanding of both particle physics, and the global properties of the universe.

\ack
RE is supported in part by the United States Department of Energy, grant DE-FG02-92ER-40704, and by an NSF Career Award PHY-0747868. 

\bibliographystyle{iopart-num}

\section*{References}
\bibliography{neutrino}

\end{document}